\documentclass[final,times,twocolumn]{elsarticle}
\usepackage{amsmath}
\usepackage{amssymb}
\usepackage{graphics}

\journal{Physics Letters A}

\begin{document}

\begin{frontmatter}

\title{Interplay between spin frustration and thermal entanglement in the exactly solved Ising-Heisenberg tetrahedral chain}  
\author[ufla]{Onofre Rojas\corref{cor2}} 
\cortext[cor2]{O. R. thanks FAPEMIG and CNPq for financial support, and Faculty of Science of P. J. \v{S}af\'{a}rik University in Ko\v{s}ice - Slovakia 
for the afforded hospitality.\\ {\it Email}: ors@dex.ufla.br (Onofre Rojas); \\Tel.: +5535 38291954}  
\author[upjs]{Jozef Stre\v{c}ka\corref{cor1}}
\cortext[cor1]{J.S. acknowledges financial support of Ministry of Education 
of Slovak Republic provided under the VEGA grant No.~1/0234/12 and by the grant 
of the Slovak Research and Dvelopment Agency under the contract No. APVV-0132-11.}
\author[ufal]{Marcelo L. Lyra} 
\address[ufla]{Departamento de Ciencias Exatas, Universidade Federal de Lavras, 
37200-000, Lavras-MG, Brazil}
\address[upjs]{Department of Theoretical Physics and Astrophysics, 
Faculty of Science, P. J. \v{S}af\'{a}rik University, Park Angelinum 9,
040 01 Ko\v{s}ice, Slovakia}
\address[ufal]{Instituto de F\'isica, Universidade Federal
de Alagoas, 57072-970, Maceio-AL, Brazil}

\begin{abstract}
The spin-1/2 Ising-Heisenberg tetrahedral chain is exactly solved using its local gauge symmetry (the total spin of the Heisenberg bonds  is locally conserved) and the transfer-matrix approach.  Exact results derived for spin-spin correlation functions are employed to obtain the frustration temperature. In addition, we have exactly calculated a concurrence quantifying thermal  entanglement. 
It is shown that the frustration and threshold temperature coincide at sufficiently low temperatures, while they  exhibit a very different behavior in the high-temperature region when tending towards completely different asymptotic limits.
 The threshold temperature additionally shows a notable reentrant behavior when it extends  over a narrow temperature region above the classical ground state without any quantum correlations. 
 \end{abstract}


\end{frontmatter}

\section{Introduction}

The term frustration was first introduced into magnetism by Toulouse \cite{toulouse}. According to 
the original concept, the frustration is basically a topological property and refers to a negative 
sign of the product of coupling constants implying an incapability of spins to satisfy all pair spin-spin 
interactions between them. One of the simplest frustrated spin systems is the Ising model 
on a triangular lattice with the unique antiferromagnetic nearest-neighbor coupling \cite{wannier}. 
For many years, frustrated spin systems have been widely studied mainly in connection with their 
unconventional low-temperature behavior and reentrant phase transitions (see Refs. \cite{liebmann,diep} and references cited therein). 
An alternative definition of the spin frustration was introduced by dos Santos and Lyra, which have proposed 
a negative sign of the product of pair correlation functions as another useful hallmark of the spin frustration 
\cite{dos Santos-Lyra}. This latter definition has an advantage in that it provides plausible criterion whether 
or not the spin system is frustrated at finite (non-zero) temperature as well.

Recently, quantum spin chains have attracted a lot of attention as they provide an useful playground for the investigation 
of thermal entanglement and moreover, they represent promising candidates for the quantum information processing \cite{amicovedral,horodecki}. 
It is worth noticing that the entanglement is one of the most exciting properties of quantum systems, which does not have any classical analog 
and is most clearly manifested at zero temperature. However, thermal fluctuations gradually destroy the quantum entanglement, 
which may finally completely disappear above a certain temperature called as the threshold temperature \cite{wang,wang-sun,su,markham,nakata}.

A theoretical investigation of the Heisenberg model on a two-leg ladder is motivated not only from the theoretical point of view, 
but also from the experimental viewpoint \cite{rice,dagotto}. Unfortunately, the frustrated Heisenberg model on a two-leg ladder cannot 
be rigorously solved quite generally and there exist only few exact results for some special cases, such as the Heisenberg tetrahedral chain 
\cite{gelfand,richter,schmidt,derzhko} or the Heisenberg ladder with the first- and second-neighbor interaction exhibiting an exact dimer ground state \cite{takano}. 
Owing to this fact, different powerful analytical and numerical approaches have been employed in order to examine the magnetic behavior of 
the frustrated Heisenberg two-leg ladder (see Refs. \cite{mila,oitmaa,honecker,high-t,shik,michaud} and references therein). On the other hand, the ground state of frustrated Ising-Heisenberg ladder can be rigorously attained from a precise mapping correspondence with the quantum Ising chain of composite spins and this model still displays typical quantum features including remarkable quantum phase transitions as discussed in detail by Verkholyak and Stre\v{c}ka \cite{ladder,taras}. Other similar exact studies of the ground state have been performed for the slightly different Ising-Heisenberg chains made up of corner sharing double-tetrahedra \cite{diana,vadim}. Recently, the magneto-caloric properties of the Ising two-leg ladder was discussed within the framework of the transfer-matrix approach \cite{yan-qi}. Another interesting quantum spin model represents the Heisenberg tetrahedral chain developed by Niggemann \textit{et al}. \cite{niggemann}, whose ground state and thermodynamic properties can be obtained from the local gauge symmetry revealing the hidden Ising symmetry. In the present work, we will exactly treat the analogous Ising-Heisenberg tetrahedral chain with the aim to examine thermal properties (partition function, free energy, specific heat and correlation functions) in addition to the ground state.  
 
The outline of this Letter is as follows. The tetrahedral Ising-Heisenberg chain is introduced in Sec. 2 together with its ground-state
phase diagram. In Sec. 3 we present basic steps of the calculation procedure suitable for evaluating finite-temperature properties
such as the correlation functions, the frustration temperature, the threshold temperature for bipartite entanglement, as well as, 
the zero-field specific heat. Finally, we will draw our conclusions and summarize the most important findings in Sec. 4.

\section{The Ising-Heisenberg tetrahedral chain}

\begin{figure}
\begin{center}
\includegraphics[width=0.45\textwidth]{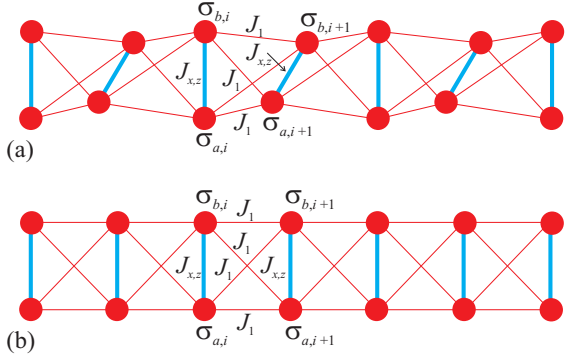}
\end{center}
\vspace{-0.6cm}
\caption{(a) A schematic representation of the spin-$\frac{1}{2}$ Ising-Heisenberg tetrahedral chain. Thick (blue) lines correspond to the Heisenberg coupling $J_{x,z}$, while thin (red) lines correspond to the Ising coupling $J_1$; (b) the tetrahedral chain can also be viewed as a two-leg ladder with equal interactions along the legs and diagonals.}
\label{fig1}
\end{figure}

Motivated by the comments given in the introduction, let us consider the spin-$\frac{1}{2}$ Ising-Heisenberg chain of edge-sharing tetrahedra as schematically illustrated in Fig.~\ref{fig1}(a). For easy reference, the investigated spin system will be hereafter referred to as the Ising-Heisenberg tetrahedral chain and it may be alternatively viewed as the particular case of the Ising-Heisenberg two-leg ladder \cite{ladder,taras} with a Heisenberg intra-rung interaction and a unique Ising interaction along the legs and diagonals, respectively (see Fig.~\ref{fig1}(b)). The Hamiltonian of the spin-$\frac{1}{2}$ Ising-Heisenberg tetrahedral chain is given by
\begin{alignat}{1}
\hat{\cal H} = &-\sum_{i=1}^{N}\left\{ J_x (\hat{\sigma}_{a,i}^{x}\hat{\sigma}_{b,i}^{x} + \hat{\sigma}_{a,i}^{y}\hat{\sigma}_{b,i}^{y}) 
                         + J_{z}\hat{\sigma}_{a,i}^{z}\hat{\sigma}_{b,i}^{z}\right.\nonumber \\
 &\left.+J_{1}(\hat{\sigma}_{a,i}^{z}+\hat{\sigma}_{b,i}^{z})(\hat{\sigma}_{a,i+1}^{z}+\hat{\sigma}_{b,i+1}^{z})\right\},
\label{eq:Ham-orig}
\end{alignat}
where $\hat{\sigma}_{a,i}^{\alpha}$ ($\hat{\sigma}_{b,i}^{\alpha}$) denote spatial components $\alpha=\{x,y,z\}$ of the spin-$\frac{1}{2}$ Pauli operator 
at a site labeled by suffix $i$ and $a$ ($b$), respectively (see Fig. \ref{fig1}). Note furthermore that the interaction terms $J_x$ and $J_{z}$ determine the spatially anisotropic XXZ Heisenberg coupling, whereas the interaction term $J_{1}$ represents the Ising coupling.

To proceed further with calculations, let us introduce three spatial components of the total spin angular momentum $\hat{S}_{i}^{\alpha}=\hat{\sigma}_{a,i}^{\alpha}+\hat{\sigma}_{b,i}^{\alpha}$ for two spins coupled by the Heisenberg interaction. Using the above definition one simply obtains the identity $(\hat{S}_{i}^{\alpha})^{2}=\frac{1}{2}+2\hat{\sigma}_{a,i}^{\alpha}\hat{\sigma}_{b,i}^{\alpha}$, which allows us to express both parts of the Heisenberg interaction in terms of two commuting operators $\boldsymbol{\hat{S}}_{i}^{2}=\boldsymbol{\hat{S}}_{i} \cdot \boldsymbol{\hat{S}}_{i}$ and $\hat{S}_{i}^{z}$
\begin{equation}
J_x \left(\hat{\sigma}_{a,i}^{x}\hat{\sigma}_{b,i}^{x}+\hat{\sigma}_{a,i}^{y}\hat{\sigma}_{b,i}^{y}\right)
=\frac{J_x}{2}\boldsymbol{\hat{S}}_{i}^{2}-\frac{J_x}{2}\left(\hat{S}_{i}^{z}\right)^{2}-\frac{J_x}{2},
\label{eq:spin-transf}
\end{equation}
and
\begin{equation}
J_{z}\hat{\sigma}_{a,i}^{z}\hat{\sigma}_{b,i}^{z}=\frac{J_{z}}{2}\left(\hat{S}_{i}^{z}\right)^{2}-\frac{J_{z}}{4}.
\label{eq:spin-transf-1}
\end{equation}
It is quite obvious from Eqs. (\ref{eq:spin-transf}) and (\ref{eq:spin-transf-1}) that the Heisenberg interaction can be rewritten with the help of total spin angular momentum $\boldsymbol{\hat{S}}_{i}^{2}$ and its $z$th spatial component $\hat{S}_{i}^{z}$, which correspond to conserved quantities with well defined quantum spin numbers $S_i (S_i + 1)$ and $S_i^z = -S_i, -S_i + 1, \ldots , S_i$, respectively. Besides, the Ising interaction as the other interaction term entering the Hamiltonian (\ref{eq:Ham-orig}) can also be trivially expressed by means of the spin operator $\hat{S}_{i}^{z}$. Bearing all this in mind, the Hamiltonian of the spin-$\frac{1}{2}$ Ising-Heisenberg tetrahedral chain can be decomposed into two parts $\hat{\cal H} = \hat{\cal H}_{0}+\hat{\cal H}'$, whereas $\hat{\cal H}_{0}=\frac{N}{4}(2J_x+J_{z})$ and
\begin{alignat}{1}
\hat{\cal H}'= & -\sum_{i=1}^{N}\left\{ J_{1} \hat{S}_{i}^{z} \hat{S}_{i+1}^{z}+\frac{J_x}{4}(\boldsymbol{\hat{S}}_{i}^{2}+\boldsymbol{\hat{S}}_{i+1}^{2})\right.\nonumber \\
& \left.+\frac{J_{z}-J_x}{4}\left[\left(\hat{S}_{i}^{z}\right)^{2}+\left(\hat{S}_{i+1}^{z}\right)^{2}\right]\right\}.
\label{eq:Ham-eff}
\end{alignat}
Up to unimportant constant term $\hat{\cal H}_{0}$, the Hamiltonian (\ref{eq:Ham-orig}) of the spin-$\frac{1}{2}$ Ising-Heisenberg tetrahedral chain can be rigorously mapped to the Hamiltonian (\ref{eq:Ham-eff}) of the classical chain of composite Ising spins, since the quantum spin number for the total angular momentum is either $S_i = 0$ or $1$. It is noteworthy that the Ising interaction $J_1$ directly determines the effective nearest-neighbor interaction in the equivalent classical chain of composite Ising spins, while the spatial anisotropy in the XXZ Heisenberg interaction determines the effective single-ion anisotropy $\frac{J_{z}-J_x}{4}$ acting on the composite spins. It should be emphasized, moreover, that the relevant gauge transformation directly brings the effective Hamiltonian (\ref{eq:Ham-eff}) into a diagonal form. Hence, one easily obtains also a diagonal form for the Hamiltonian of the spin-$\frac{1}{2}$ Ising-Heisenberg tetrahedral chain  
\begin{alignat}{1}
\cal H = & \frac{N}{4}(2J_x+J_{z}) -\sum_{i=1}^{N}\left\{ \frac{J_x}{4} \left[S_i(S_i + 1) + S_{i+1}(S_{i+1} + 1) \right]\right.\nonumber \\
& \left.+J_{1} S_{i}^{z}S_{i+1}^{z}+\frac{J_{z}-J_x}{4}\left[\left(S_{i}^{z}\right)^{2}+\left(S_{i+1}^{z}\right)^{2}\right]\right\}.
\label{eq:diagham}
\end{alignat}

It is worthwhile to remark that the transformation between the quantum tetrahedral spin chain and the classical chain of composite spins is only possible because the total spin of the Heisenberg bonds is locally conserved quantity for the Hamiltonian \eqref{eq:Ham-orig} whenever the Ising interactions along the legs and diagonals are identical. If both the Ising interactions would be different, the total spin of the Heisenberg bonds does not represent the locally conserved quantity and this would consequently prevent a rigorous mapping onto the classical chain of composite spins even though the investigated spin system could still be mapped onto the quantum Ising chain of composite spins \cite{ladder,taras}. However, the latter quantum chain of composite spins can exactly be treated only at zero temperature unlike the former classical chain of composite spins (see Sec. 3). It should be also noted here that the Heisenberg tetrahedral chain under the local gauge symmetry has been studied by Niggeman \textit{et al}. \cite{niggemann} when combining the fragmentation approach with numerical calculations, while the relevant high-temperature expansion has been developed in Ref.~\cite{high-t}.

\subsection{Ground-state phase diagram}

\begin{figure}
\begin{center}
\includegraphics[width=0.35\textwidth]{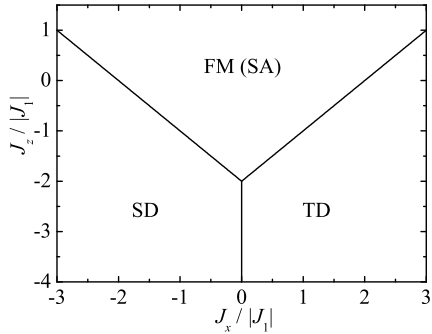}
\end{center}
\vspace{-0.8cm}
\caption{The ground-state phase diagram in the $J_x/|J_{1}|-J_{z}/|J_{1}|$ plane. 
An upper sector of the phase diagram contains FM (SA) ground state for the particular 
case with $J_{1}>0$ ($J_{1}<0$).}
\label{fig:gs}
\end{figure}

The diagonalized form of the Hamiltonian (\ref{eq:diagham}) can be straightforwardly used in order to obtain all possible ground states of the spin-$\frac{1}{2}$ Ising-Heisenberg tetrahedral chain. The Hamiltonian (\ref{eq:diagham}) implies an existence of four different ground states, namely, the ferromagnetic state (FM), 
the superantiferromagnetic state (SA), the triplet-dimer state (TD), and the singlet-dimer state (SD), which are unambiguously given by the following eigenvectors
\begin{align}
|\mbox{FM}\rangle= & \prod_{i=1}^{N}|\begin{smallmatrix}+\\
+
\end{smallmatrix}\rangle_{i},\label{fm}\\
|\mbox{SA}\rangle= & \prod_{i=1}^{N/2}|\begin{smallmatrix}+\\
+
\end{smallmatrix}\rangle_{2i-1}\otimes|\begin{smallmatrix}-\\
-
\end{smallmatrix}\rangle_{2i},\label{sa}\\
|\mbox{TD}\rangle= & \prod_{i=1}^{N}\frac{1}{\sqrt{2}}\left(|\begin{smallmatrix}+\\
-
\end{smallmatrix}\rangle_i+|\begin{smallmatrix}-\\
+
\end{smallmatrix}\rangle_i\right), \label{td}\\
|\mbox{SD}\rangle= & \prod_{i=1}^{N}\frac{1}{\sqrt{2}}\left(|\begin{smallmatrix}+\\
-
\end{smallmatrix}\rangle_i-|\begin{smallmatrix}-\\
+
\end{smallmatrix}\rangle_i\right).
\label{sd}
\end{align}
In above, the upper $\pm$ sign applies for $\sigma_{a,i}^z = \pm \frac{1}{2}$ and the lower $\pm$ sign stands for $\sigma_{b,i}^z = \pm \frac{1}{2}$. 
The corresponding energy per unit cell for each individual ground state reads
\begin{align}
E_{\mbox{FM}}= & - \frac{J_{z}}{4}- J_{1}, \nonumber \\
E_{\mbox{SA}}= & - \frac{J_{z}}{4}+ J_{1}, \nonumber \\
E_{\mbox{TD}}= &   \frac{J_{z}}{4}- \frac{J_x}{2}, \nonumber \\
E_{\mbox{SD}}= &   \frac{J_{z}}{4}+ \frac{J_x}{2}.
\label{ee}
\end{align}
By making use of the eigenenergies (\ref{ee}) we have constructed the zero-temperature phase diagram, which is displayed in Fig. \ref{fig:gs} 
in the $J_x/|J_{1}|-J_{z}/|J_{1}|$ plane. The disordered SD state with singlet dimers placed on all Heisenberg bonds becomes the ground state 
whenever $J_x<0$ and $J_{z}<-J_x - 2|J_{1}|$. On the other hand, the similar disordered TD state is being the respective ground state if 
$J_x>0$ and $J_{z} < J_x - 2|J_{1}|$. The most fundamental difference between TD and SD states lies in the symmetric versus antisymmetric 
quantum superposition of two antiferromagnetic states as evidenced by the eigenvectors \eqref{td} and \eqref{sd}, respectively. Note furthermore that TD and SD 
ground states are of a purely quantum nature, whereas  they become the relevant ground states only if the intra-rung Heisenberg interaction  is sufficiently strong. 
Finally, the less interesting classical FM (SA) state without any quantum correlations constitutes the ground state for $J_1>0$ ($J_1<0$) 
on assumption that $J_{z} > |J_x| - 2 |J_{1}|$.

\section{Thermodynamics}

In this part, let us calculate finite-temperature properties of the spin-$\frac{1}{2}$ Ising-Heisenberg tetrahedral chain. To gain the partition function, 
it is quite convenient to adopt a trace invariance and to calculate the partition function using the diagonalized form of the Hamiltonian (\ref{eq:diagham})  
\begin{equation}
\mathcal{Z}_N = \mathrm{Tr} \, \mathrm{e}^{-\beta \hat{\cal H}} = \mathrm{Tr} \, \mathrm{e}^{-\beta {\cal H}}.
\label{pf}
\end{equation}
Here, $\beta=1/(k_{\rm B} T)$, $k_{\rm B}$ is being the Boltzmann's constant and $T$ is the absolute temperature. Let us first rewrite the diagonalized form of the Hamiltonian (\ref{eq:diagham}) into the following compact form
\begin{alignat}{1}
{\cal H} = {\cal H}_0 + \sum_{i=1}^{N} {\cal H}_i, 
\label{eq:diagham1}
\end{alignat}
with the Hamiltonian ${\cal H}_i$ defined as 
\begin{alignat}{1}
{\cal H}_i = & - J_{1} S_{i}^{z}S_{i+1}^{z} - \frac{J_{z}-J_x}{4}\left[\left(S_{i}^{z}\right)^{2}+\left(S_{i+1}^{z}\right)^{2}\right] \nonumber \\
& - \frac{J_x}{4} \left[S_i(S_i + 1) + S_{i+1}(S_{i+1} + 1) \right].
\label{eq:diagham2}
\end{alignat}
Then, the partition function can be partially factorized if substituting Eq.~\eqref{eq:diagham1} into (\ref{pf})  
\begin{equation}
\mathcal{Z}_N = \mathrm{e}^{-\beta {\cal H}_0} \mathrm{Tr} \prod_{i=1}^{N} \mathrm{e}^{-\beta {\cal H}_i} 
            = \mathrm{e}^{-\beta {\cal H}_0} \mathrm{Tr} \prod_{i=1}^{N} \boldsymbol{T} (S_i, S_i^z, S_{i+1}, S_{i+1}^z)
\label{pff}
\end{equation}
and the latter expression behind the product symbol can be subsequently identified as the usual transfer matrix 
\begin{align}
\boldsymbol{T} (S_i, S_i^z&, S_{i+1}, S_{i+1}^z) = \langle S_i, S_i^z| \mathrm{e}^{-\beta {\cal H}_i} | S_{i+1}, S_{i+1}^z \rangle \nonumber \\
&= \left(\begin{array}{cccc}
{x}^{2}{y}^{2} z & {x}^{3}y & xy & {x}^{2}{y}^{2}z^{-1}\\
{x}^{3}y & {x}^{4} & {x}^{2} & {x}^{3}y\\
xy & {x}^{2} & 1 & xy\\
{x}^{2}{y}^{2}z^{-1} & {x}^{3}y & xy & {x}^{2}{y}^{2}z
\end{array}\right)
\label{tm}
\end{align}
with $x=\mathrm{e}^{\beta J_x/4}$, $y=\mathrm{e}^{\beta J_{z}/4}$, and $z=\mathrm{e}^{\beta J_{1}}$.
Hence, it follows that the partition function can be evaluated by the standard transfer-matrix approach \cite{baxter}.
Moreover, the eigenvalue problem for the transfer matrix \eqref{tm} can easily be 
solved from the secular determinant, which leads to the following quartic equation
\begin{equation}
\lambda\left(\lambda^{2}-a_{1}\lambda+a_{0}\right)\left[\lambda-x^{2}y^{2}(z-z^{-1})\right]=0
\label{eq:cubic-eq}
\end{equation}
with 
\begin{align}
a_{0}= & x^{2}y^{2}z^{-1}(z-1)^{2}(x^{4}+1),\label{eq:a0}\\
a_{1}= & x^{4}+1+x^{2}y^{2}(z+z^{-1}).\label{eq:a1}
\end{align}

After a straightforward calculation one finds all four transfer-matrix eigenvalues 
\begin{align*}
\lambda_{0}= & \tfrac{1}{2}\left(a_{1}+\sqrt{a_{1}^{2}-4a_{0}}\right),\\
\lambda_{1}= & \tfrac{1}{2}\left(a_{1}-\sqrt{a_{1}^{2}-4a_{0}}\right),\\
\lambda_{2}= & x^{\text{2}}y^{2}\left(z-z^{-1}\right),\\
\lambda_{3}= & 0,
\end{align*}
whereas the largest eigenvalue is $\lambda_{0}$ by inspection. In the thermodynamic limit,
the free energy per unit cell is simply given by the largest eigenvalue of the transfer matrix 
\begin{equation}
f = - \beta^{-1} \lim_{N \to \infty} \frac{1}{N} \ln {\cal Z} _N= \frac{2J_x+J_{z}}{4}-\frac{1}{2\beta} \ln \lambda_{0}. 
\end{equation}
It is worth mentioning that the first constant term of the free energy was obtained during the local gauge transformation 
and this term is irrelevant for most thermodynamic quantities. After a straightforward but a little bit cumbersome manipulation, 
one may derive the following final formula for the free energy per unit cell 
\begin{equation}
f = - \beta^{-1} \ln \left[\mathrm{e}^{\frac{\beta J_z}{4}} \! \cosh (\beta J_1) + \mathrm{e}^{-\frac{\beta J_z}{4}} \! \cosh \left(\frac{\beta J_x}{2} \right) + Q \right],
\label{free}
\end{equation}
where the coefficient $Q$ is defined as follows
\begin{equation}
Q = \sqrt{\left[\mathrm{e}^{\frac{\beta J_z}{4}} \! \cosh (\beta J_1) - \mathrm{e}^{-\frac{\beta J_z}{4}} \! \cosh \left(\frac{\beta J_x}{2} \right) \right]^2 
\!\!\!+ 4 \cosh \left(\frac{\beta J_x}{2} \right)}.
\end{equation}

\subsection{Correlation functions}

Now, let us calculate a few pair correlation functions, which might be helpful in characterizing the magnetic behavior at finite temperatures.
The short-range correlation functions can in turn be calculated by differentiating the free energy \eqref{free} with respect to the relevant coupling constant. 
For instance, both spatial components of the correlation function between two spins coupled by the Heisenberg interaction can be obtained 
by differentiating the free energy \eqref{free} with respect to the relevant spatial component ($J_z$ or $J_x$) of the Heisenberg interaction 
\begin{equation}
\langle \hat{\sigma}_{a,i}^{z} \hat{\sigma}_{b,i}^{z}\rangle \!=\! \frac{1}{4Q} \! \left[\mathrm{e}^{\frac{\beta J_z}{4}} \! \cosh (\beta J_1) 
- \mathrm{e}^{-\frac{\beta J_z}{4}} \! \cosh \left(\frac{\beta J_x}{2} \right) \right]\! 
\label{zz}
\end{equation}
and
\begin{align}
& \langle \hat{\sigma}_{a,i}^{x} \hat{\sigma}_{b,i}^{x}\rangle = \langle \hat{\sigma}_{a,i}^{y} \hat{\sigma}_{b,i}^{y}\rangle 
\nonumber \\
&= \frac{\sinh \left(\frac{\beta J_x}{2} \right) \left[2 - \cosh (\beta J_1) + \mathrm{e}^{-\frac{\beta J_z}{2}} \! \cosh \left(\frac{\beta J_x}{2} \right) 
+ \mathrm{e}^{-\frac{\beta J_z}{4}} \! Q \right]}{4Q \left[\mathrm{e}^{\frac{\beta J_z}{4}} \cosh (\beta J_1) + \mathrm{e}^{-\frac{\beta J_z}{4}} 
\! \cosh \left(\frac{\beta J_x}{2} \right) + Q \right]}.
\label{xx}
\end{align}
Due to the locally conserved character of the total spin of the Heisenberg bonds, the longitudinal pair correlation functions between the nearest-neighbor spins interacting via the Ising interaction must be equal to each other. Thus, the differentiation of the free energy \eqref{free} with respect to the Ising interaction $J_1$ gives us another important pair correlation functions
\begin{align}
& \langle \hat{\sigma}_{a,i}^{z} \hat{\sigma}_{a,i+1}^{z}\rangle = \langle \hat{\sigma}_{a,i}^{z}\hat{\sigma}_{b,i+1}^{z}\rangle = \langle \hat{\sigma}_{b,i}^{z}\hat{\sigma}_{a,i+1}^{z}\rangle = \langle \hat{\sigma}_{b,i}^{z}\hat{\sigma}_{b,i+1}^{z}\rangle \nonumber \\
& = \frac{\mathrm{e}^{\frac{\beta J_z}{4}} \sinh (\beta J_1) \left[\mathrm{e}^{\frac{\beta J_z}{4}} \cosh (\beta J_1) 
        - \mathrm{e}^{-\frac{\beta J_z}{4}} \! \cosh \left(\frac{\beta J_x}{2} \right) + Q \right]}  
         {4Q \left[\mathrm{e}^{\frac{\beta J_z}{4}} \cosh (\beta J_1) + \mathrm{e}^{-\frac{\beta J_z}{4}} \! \cosh \left(\frac{\beta J_x}{2} \right) + Q \right]}. 
\label{izz}
\end{align}

\begin{figure}
\begin{center}
\hspace{-0.2cm}
\includegraphics[width=0.245\textwidth]{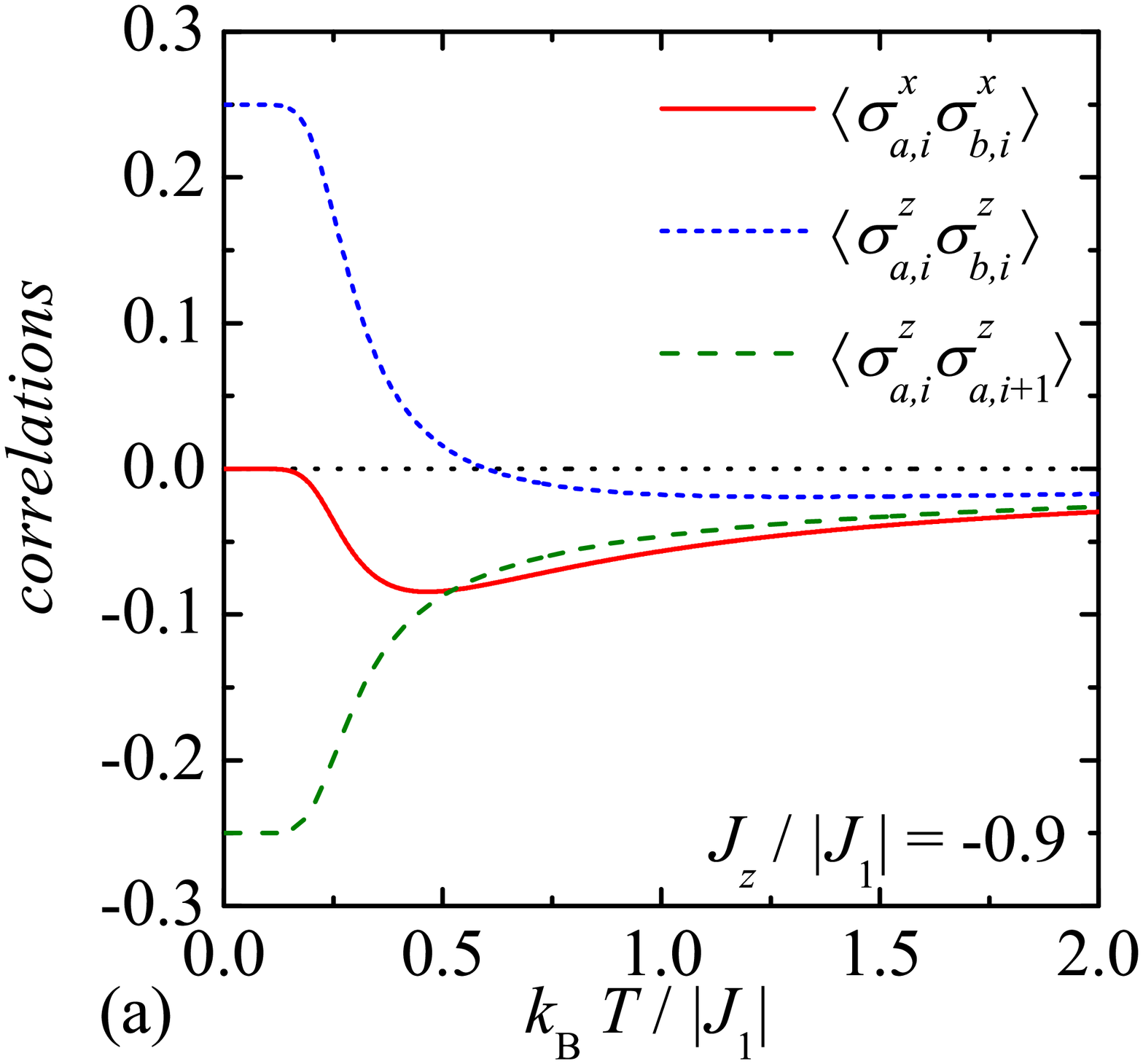}
\hspace{-0.1cm}
\includegraphics[width=0.245\textwidth]{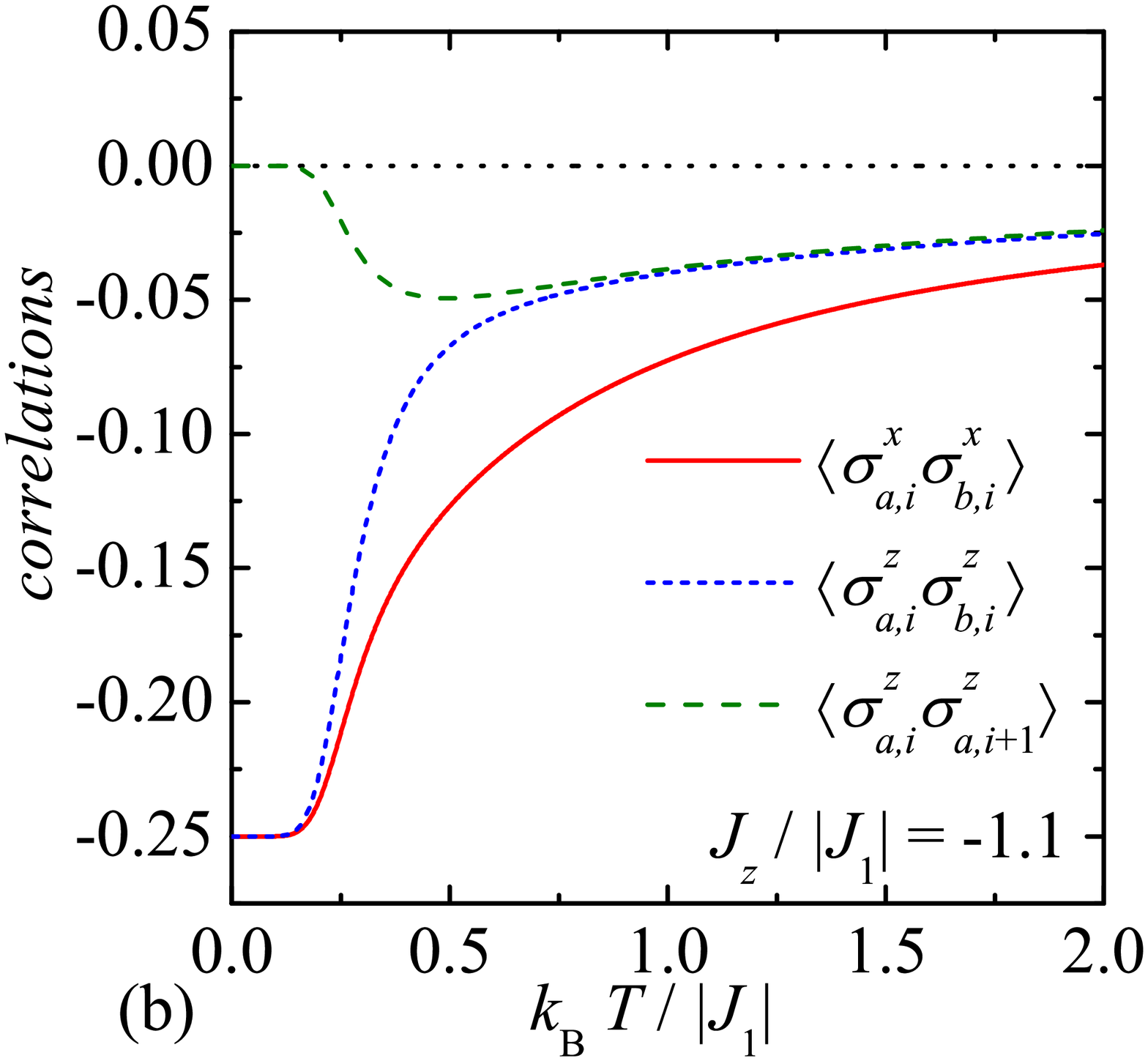}
\end{center}
\vspace{-0.7cm}
\caption{Thermal variations of the pair correlation functions for the isotropic antiferromagnetic Heisenberg interaction $J_x = J_z<0$, the antiferromagnetic Ising interaction $J_1 <0$, and two different values of the interaction ratio: (a) $J_{z}/|J_{1}|=-0.9$; (b) $J_{z}/|J_{1}|=-1.1$.}
\label{fig:Corrl}
\end{figure}

Let us analyze in detail typical temperature dependences of all calculated pair correlation functions for the most interesting particular case 
with the antiferromagnetic Ising and Heisenberg interactions. For illustration, the correlation functions are plotted in Fig.~\ref{fig:Corrl} against temperature for the isotropic Heisenberg interaction and two different values of the interaction ratio $J_{z}/|J_{1}|=-0.9$ and $-1.1$, respectively. Two selected values of the interaction ratio $J_{z}/|J_{1}|$ has been chosen so as to achieve two different ground states. As a matter of fact, the zero-temperature limits of the displayed correlation functions are consistent with a presence of the classical SA ground state in the former case with $J_{z}/|J_{1}|=-0.9$, while the existence of the more striking quantum SD ground state is evident in the latter case with $J_{z}/|J_{1}|=-1.1$. It is quite obvious from Fig.~\ref{fig:Corrl}(a) that thermal fluctuations may induce quantum correlations above the classical SA ground state if the interaction ratio $J_{z}/|J_{1}|$ is selected sufficiently close to the ground-state boundary with the SD state. It actually turns out that the transverse pair correlation function $\langle \hat{\sigma}_{a,i}^{x} \hat{\sigma}_{b,i}^{x}\rangle$ becomes non-zero upon increasing temperature and the longitudinal pair correlation function $\langle \hat{\sigma}_{a,i}^{z} \hat{\sigma}_{b,i}^{z}\rangle$ changes its signal at some finite temperature $T_{\rm f}$ called as the frustration temperature. Following the definition by dos Santos and Lyra \cite{dos Santos-Lyra}, the tetrahedral chain is frustrated above the frustration temperature as the product $\langle \hat{\sigma}_{a,i}^{z} \hat{\sigma}_{b,i}^{z}\rangle \langle \hat{\sigma}_{a,i}^{z} \hat{\sigma}_{a,i+1}^{z}\rangle \langle \hat{\sigma}_{a,i+1}^{z} \hat{\sigma}_{b,i}^{z}\rangle$ between three correlation functions along the triangular face of an elementary tetrahedron becomes negative. It is worthy of notice, moreover, that the correlation functions $\langle \hat{\sigma}_{a,i}^{z} \hat{\sigma}_{b,i}^{z}\rangle$ and $\langle \hat{\sigma}_{a,i}^{z} \hat{\sigma}_{a,i+1}^{z}\rangle$ remain negative for arbitrary temperature whenever the SD state constitutes the ground state. This result is taken to mean that the frustration persists for all temperatures 
if the SD state constitutes the ground state.  

\subsection{Thermal entanglement}

Another interesting aspect of the spin-$\frac{1}{2}$ Ising-Heisenberg tetrahedral chain surely represents a study of thermal entanglement 
with regard to its unusual quantum ground states. For this purpose, one may take advantage of the quantity referred to as the \textit{concurrence} \cite{wooters,amico}, 
which may serve as a measure of the bipartite entanglement. The concurrence is defined in terms of the reduced density matrix $\rho$ 
\begin{equation}
\mathcal{C}(\rho) = \max\{{0,2\lambda_{max}-\text{{Tr}} \, R}\},
\end{equation}
where $\lambda_{max}$ is the largest eigenvalue of the matrix
\begin{equation}
R = \sqrt{\rho \hat{\sigma}^{y} \otimes \hat{\sigma}^{y} \rho^{*}\hat{\sigma}^{y} \otimes \hat{\sigma}^{y}}.
\end{equation}
Here, $\rho^{*}$ is the complex conjugate of the reduced density matrix $\rho$ and $\hat{\sigma}^{y}$ is being the relevant Pauli matrix.
It is quite well established that matrix elements of the reduced density operator can be expressed through the correlation functions \cite{amico,bukman} 
and hence, the concurrence can be directly calculated from the correlation functions as evidenced by Amico and co-workers \cite{amico}. In our case, 
the concurrence quantifying the bipartite entanglement between two spins coupled by the Heisenberg interaction then follows from the relation
\begin{equation}
\mathcal{C} = \max \left\{0, 4 \left|\langle \hat{\sigma}_{a,i}^{x} \hat{\sigma}_{b,i}^{x} \rangle \right| 
- \left|\frac{1}{2} + 2 \langle \hat{\sigma}_{a,i}^{z} \hat{\sigma}_{b,i}^{z} \rangle \right| \right\},
\label{eq:Cnr}
\end{equation}
which just involves two pair correlation functions given by Eqs. \eqref{zz} and \eqref{xx}. It is noteworthy that an equivalent result 
could be obtained by employing the approach developed in Ref. \cite{spra}.

\begin{figure}
\begin{center}
\hspace{-0.1cm}
\includegraphics[width=0.24\textwidth]{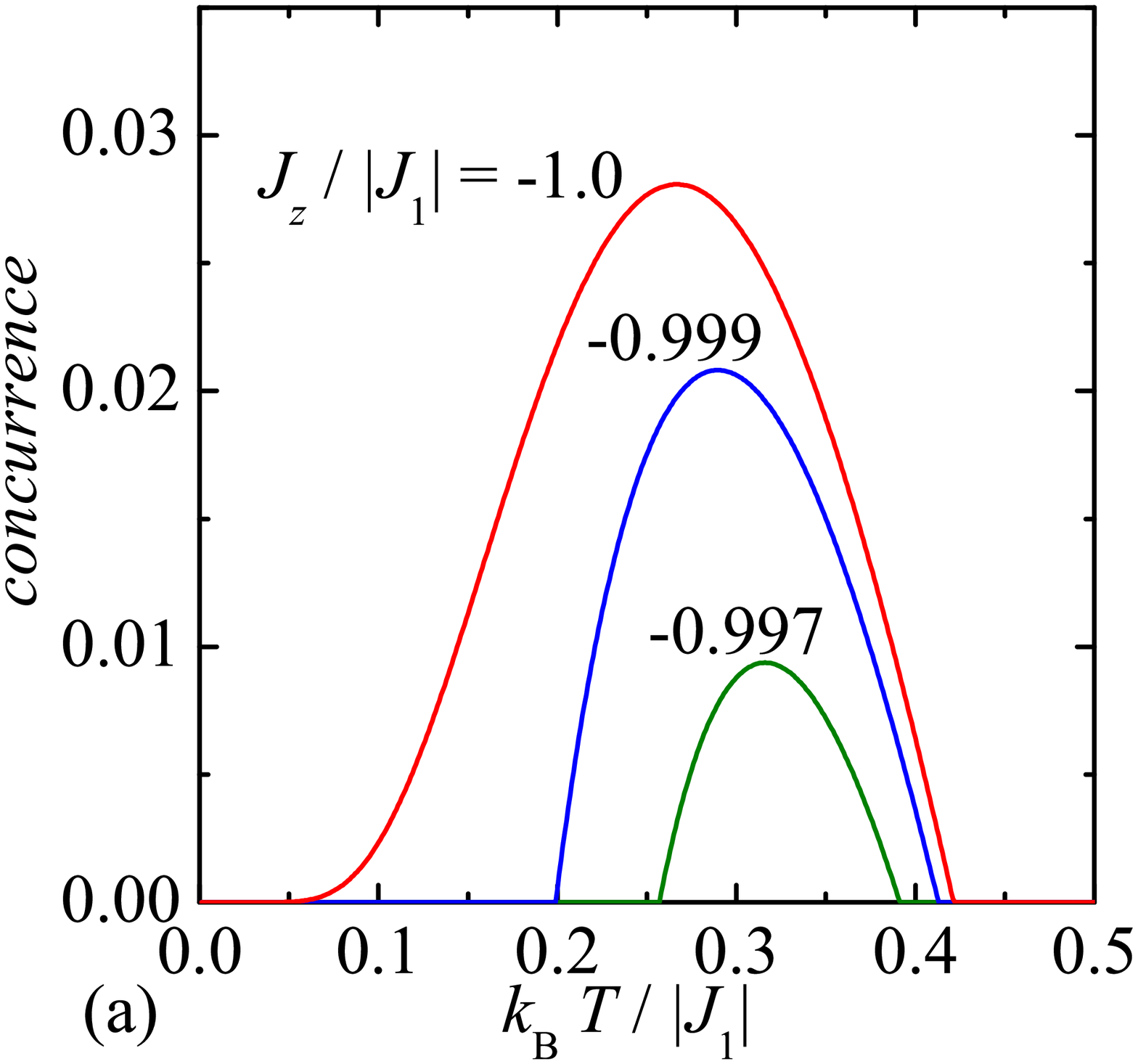}
\hspace{-0.1cm}
\includegraphics[width=0.24\textwidth]{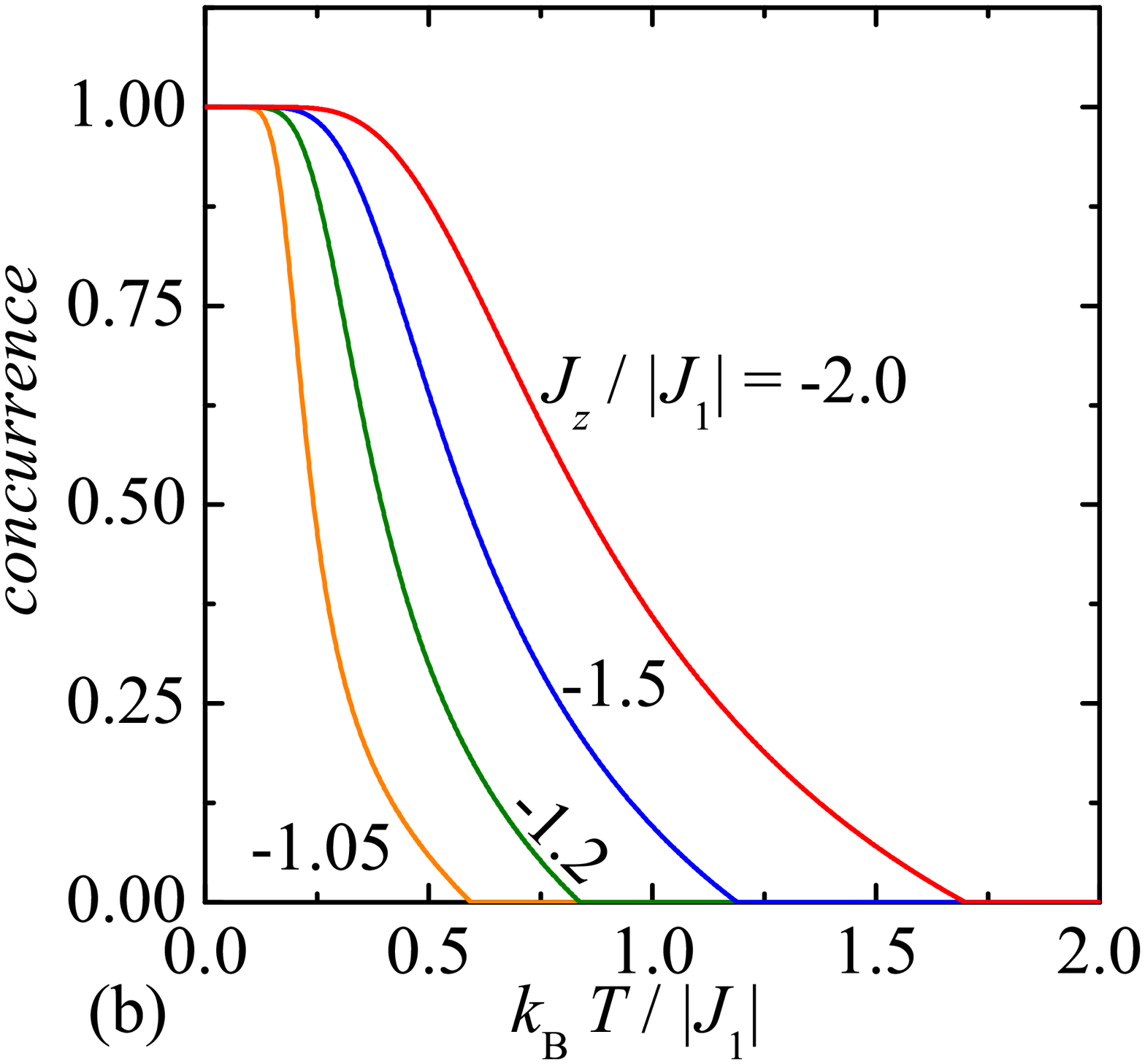}
\end{center}
\vspace{-0.7cm}
\caption{Temperature dependences of the concurrence for the isotropic antiferromagnetic Heisenberg interaction $J_x = J_z < 0$, 
the antiferromagnetic Ising interaction $J_1 <0$, and several values of the interaction ratio: (a) $J_{z}/|J_{1}| \geq -1.0$; 
(b) $J_{z}/|J_{1}| < -1.0$.}
\label{fig:Concrr}
\end{figure}

The concurrence is plotted against temperature in Fig.~\ref{fig:Concrr}(a) for a specific choice of the antiferromagnetic Ising and Heisenberg interactions, which ensures occurrence of the classical SA spin alignment in the ground state. It surprisingly turns out that the Ising-Heisenberg tetrahedral chain exhibits a weak thermal entanglement in spite of the classical SA ground state whenever a relative strength between the Heisenberg and Ising couplings is sufficiently close to the phase boundary 
$J_{z}^{\rm b}/|J_{1}|$ with the other possible SD ground state ($J_{z}^{\rm b}/|J_{1}|=-1.0$ if assuming the isotropic Heisenberg interaction $J_{x} = J_{z}$). Accordingly, the concurrence shows a striking reentrant behavior when it becomes non-zero at a lower threshold temperature and disappears at an upper threshold temperature $k_{\rm B} T_{\rm t}/|J_{1}| \approx 0.4$, whereas the lower threshold temperature gradually tends towards zero as the interaction ratio approaches the boundary value $J_{z}^{\rm b}/|J_{1}|$. It can be readily understood that the observed weak thermal entanglement comes from a temperature-induced strengthening of the transverse correlation function $\langle \hat{\sigma}_{a,i}^{x} \hat{\sigma}_{b,i}^{x}\rangle$, whose effect is also supported by a relevant crossover of the longitudinal correlation function $\langle \hat{\sigma}_{a,i}^{z} \hat{\sigma}_{b,i}^{z}\rangle$ from the ferromagnetic regime to antiferromagnetic one after passing through the frustration temperature. 

In addition, Fig.~\ref{fig:Concrr}(b) displays more typical temperature variations of the concurrence on assumption that the interaction ratio between the Heisenberg and Ising interactions supports the SD ground state. The spins coupled by the Heisenberg interaction are consequently maximally entangled at zero temperature and their quantum entanglement is gradually suppressed by thermal fluctuations. It appears worthwhile to remark that the threshold temperature above which the relevant spins become disentangled becomes higher as the magnitude of the ratio $J_{z}/|J_{1}|$ between the Heisenberg and Ising interactions increases. 

\subsection{Threshold versus frustration temperature}

Now, it might be quite interesting to compare the \textit{threshold} temperature, delimiting the regimes of entangled spins (finite concurrence) and unentangled spins (vanishing concurrence), with the \textit{frustration} temperature that delimits the regimes with (non frustrated) and without (frustrated) a well defined spin configuration among the three spins forming a triangular face of an elementary tetrahedron. It is quite obvious from our previous analysis of the correlation functions that the absence of spin correlations can be directly connected with a disappearance of the longitudinal pair correlation function $\langle \hat{\sigma}_{a,i}^{z} \hat{\sigma}_{b,i}^{z}\rangle$ between two spins coupled by the Heisenberg interaction. While the threshold temperature can be calculated from the condition $\mathcal{C} = 0$ by making use of Eq.~\eqref{eq:Cnr} for the concurrence, the frustration temperature can be either calculated from the condition $\langle \hat{\sigma}_{a,i}^{z} \hat{\sigma}_{b,i}^{z}\rangle \langle \hat{\sigma}_{a,i}^{z} \hat{\sigma}_{a,i+1}^{z}\rangle \langle \hat{\sigma}_{a,i+1}^{z} \hat{\sigma}_{b,i}^{z}\rangle = 0$ or the simpler equivalent condition $\langle \hat{\sigma}_{a,i}^{z} \hat{\sigma}_{b,i}^{z}\rangle = 0$ after taking into account Eq.~\eqref{zz} for the relevant correlation function. 

\begin{figure}
\begin{center}
\includegraphics[width=0.35\textwidth]{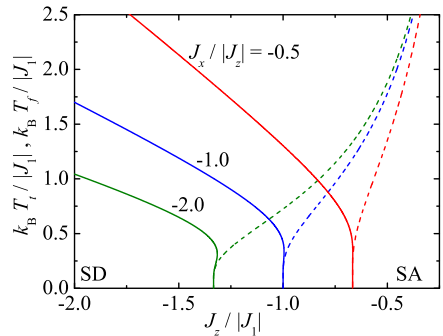}
\end{center}
\vspace{-0.8cm}
\caption{The threshold temperature (solid lines) and frustration temperature (broken lines) as a function of the interaction ratio $J_{z}/|J_{1}|$ 
for the antiferromagnetic Ising ($J_1<0$) interaction and three different values of the exchange anisotropy $J_x/|J_z|$ = -0.5, -1.0 and -2.0.}
\label{fig:Frustration-T}
\end{figure}

The frustration and threshold temperatures are plotted in Fig.~\ref{fig:Frustration-T} against the interaction ratio $J_{z}/|J_{1}|$ for three different values of the exchange anisotropy $J_x/|J_z|$ of the antiferromagnetic Heisenberg interaction. It is worthwhile to remark that the Ising-Heisenberg tetrahedral chain is thermally entangled just in the parameter region below the line of threshold temperatures (solid lines), which shows a relatively narrow reentrant behavior in a close vicinity of the ground-state boundary between the SD and SA ground states in agreement with our previous analysis of the concurrence. On the other hand, the Ising-Heisenberg tetrahedral chain is affected by a spin frustration only above the line of frustration temperatures (broken lines), which also starts from the ground-state boundary between the SD and SA ground states but asymptotically tends towards infinity as $J_{z}/|J_{1}| \rightarrow 0$. Altogether, the frustration and threshold temperature remarkably coincide at sufficiently low temperatures, while they exhibit a very different behavior in the high-temperature region when tending towards completely different asymptotic limits. It is quite interesting to notice, moreover, that the thermally entangled region above the quantum SD ground state as well as the more peculiar thermally entangled region above the classical SA ground state fall into the frustrated region. Hence, it could be concluded that the spin frustration seems to be an essential ingredient, which is needed in order to produce the thermal entanglement in the antiferromagnetic Ising-Heisenberg tetrahedral chain. It is worthy to mention, moreover, that the relevant behavior of the ferromagnetic Ising-Heisenberg tetrahedral chain has no special features at least for the isotropic case with $J_z=J_x>0$, because the ground state is then always classical and there will be neither frustration nor threshold temperature. In addition, the other particular case with $J_1>0$ and $J_x=-J_z>0$ shows a completely analogous behavior as the investigated antiferromagnetic tetrahedral chain with $J_1<0$ and $J_x=J_z<0$ due to the relevant symmetry of the interactions $J_1$ and $J_x$ (see Eq.~\eqref{zz}).

\subsection{Specific heat}

\begin{figure}
\begin{center}
\hspace{-0.1cm}
\includegraphics[width=0.24\textwidth]{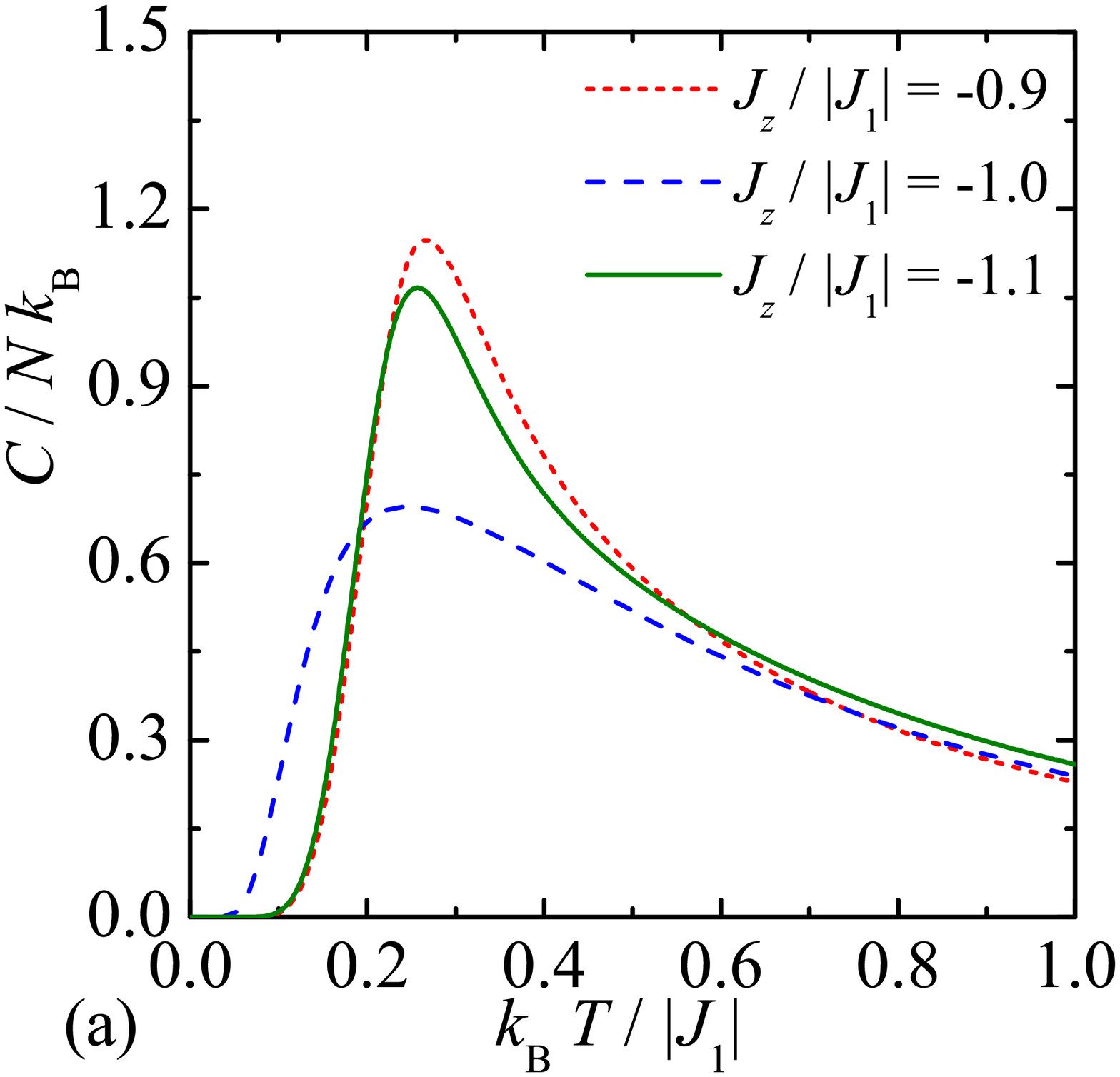}
\hspace{-0.1cm}
\includegraphics[width=0.24\textwidth]{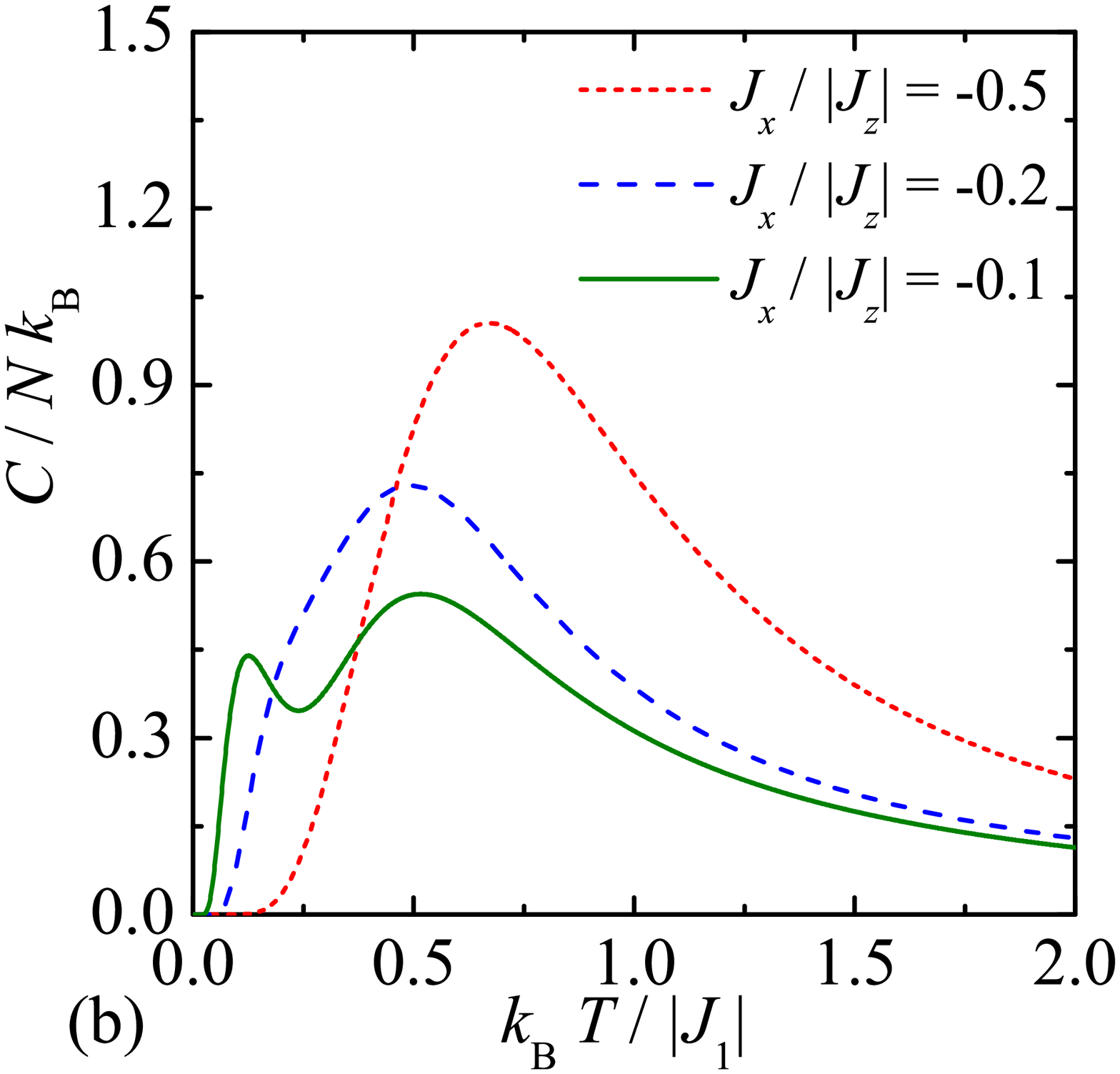}
\end{center}
\vspace{-0.7cm}
\caption{(a) Thermal variations of the specific heat for the isotropic antiferromagnetic Heisenberg interaction $J_x = J_z < 0$ by selecting the interaction ratio $J_{z}/|J_{1}|$ at and close to the ground-state boundary between the SA and SD ground states; (b) Thermal variations of the specific heat for the fixed value of the interaction ratio $J_{z}/|J_{1}| = -3$ and three different values of the exchange anisotropy $J_x/|J_z|$ in the Heisenberg interaction.}
\label{fig:Specific-heat}
\end{figure}

Last but not least, let us turn our attention to a discussion of typical temperature dependences of the specific heat numerically calculated from the free energy \eqref{free} with the help of thermodynamic relation $C= - N T\frac{\partial^{2}f}{\partial T^{2}}$. Fig.~\ref{fig:Specific-heat}(a) displays typical thermal variations of the specific heat for the isotropic antiferromagnetic Heisenberg coupling $J_x = J_z < 0$ by choosing the interaction ratio $J_{z}/|J_{1}|$ close to and precisely at the ground-state boundary between the SA and SD ground states. Under this condition, the specific heat generally exhibits the temperature dependence with a single round Schottky-type maximum irrespective of a relative strength $J_{z}/|J_{1}|$ between both the interaction parameters, whereas there does not appear any additional low-temperature maximum that would reflect thermal excitations between the ground state and the respective low-lying excited state (either from SA to SD state or vice versa). Contrary to this, the more diverse temperature variations of the heat capacity can be observed by considering the more general case of the anisotropic Heisenberg coupling (see Fig.~\ref{fig:Specific-heat}(b)). In addition to the standard thermal dependences of the specific heat with a single more or less symmetric round maximum, one may also detect a more pronounced double-peak temperature dependence as the exchange anisotropy strengthens (i.e. $J_{x}/|J_z| \to 0$). It should be pointed out that the Ising-like exchange anisotropy drives the investigated spin system close to the ground-state boundary between two quantum SD and TD ground states and thus, the additional low-temperature peak clearly reflects the low-lying thermal excitation from the SD ground state towards the excited TD state.

\section{Conclusion}

In this work, the spin-$\frac{1}{2}$ Ising-Heisenberg tetrahedral chain has been exactly solved by employing its local gauge symmetry and the transfer-matrix technique.
It is worthwhile to recall that the investigated spin system can be alternatively viewed as the spin-$\frac{1}{2}$ Ising-Heisenberg two-leg ladder with equal Ising interactions along the legs and diagonals, respectively. The elaborated rigorous procedure enabled us to obtain exact results for basic thermodynamic quantities (e.g. free energy, specific heat) and several pair correlation functions, which were subsequently used for a calculation of the concurrence quantifying the thermal entanglement between two spins coupled 
by the Heisenberg interaction, the frustration and threshold temperature. It has been demonstrated that the frustration and threshold temperature coincide at low enough temperatures, while they tend towards completely different asymptotic limits in the high-temperature range. It actually turns out that the frustration temperature diverges 
as the interaction ratio $J_{z}/|J_{1}| \rightarrow 0$, while the threshold temperature goes to infinity if $J_{z}/|J_{1}| \rightarrow -\infty$.  An interesting observation is that the investigated Ising-Heisenberg tetrahedral chain is thermally entangled just in the frustrated region and 
hence, the frustration seems to be an essential ingredient in order to produce the quantum entanglement. It is worth noticing that the threshold temperature for the pairwise entanglement exhibits a striking reentrant behavior. Therefore, it can emerge at moderate temperatures above the classical ground state without any quantum correlations. This feature unveils that the usual view point of thermal fluctuations as a mechanism through it quantum correlations are destroyed is not generally correct. In systems with an ordered classical ground state but close to a frustration point, thermal fluctuations can firstly destroy the magnetic local order, thus leaving space to the emergence of quantum entanglement at finite temperatures. Only at sufficiently high temperatures, the absence of quantum entanglement is recovered.
Besides, we have also investigated in particular typical temperature variations of the specific heat, which may exhibit a thermal dependence with (without) an anomalous low-temperature peak for a relatively strong (weak) Heisenberg interaction.
Finally, it should be also mentioned that our rigorous procedure can be adapted in a rather straightforward way to account for the non-zero external magnetic field as well. 
Our future work will continue in this direction.




\begin{thebibliography}{50}

\bibitem{toulouse}
G. Toulouse, Commun. Phys. \textbf{2} (1977) 115.

\bibitem{wannier}
G. H. Wannier, Phys. Rev. \textbf{79} (1950) 357; Erratum: Phys. Rev. B \textbf{7} (1973) 5017.

\bibitem{liebmann}
R. Liebmann, Statistical Mechanics of Periodic Frustrated Ising Systems
(Springer-Verlag, Berlin, 1986). 

\bibitem{diep}
H. T. Diep, H. Giacomini, in Frustrated Spin Systems, edited
by H. T. Diep (World Scientific, Singapore, 2004).

\bibitem{dos Santos-Lyra}
R. J. V. dos Santos, M. L. Lyra, Physica A \textbf{182} (1992) 133.

\bibitem{amicovedral}
L. Amico, R. Fazio, A. Osterloh, V. Vedral, Rev. Mod. Phys. \textbf{80} (2008) 517. 

\bibitem{horodecki}
R. Horodecki, P. Horodecki, M. Horodecki, K. Horodecki, Rev. Mod. Phys. \textbf{81} (2009) 865. 

\bibitem{wang}
X. Wang, Phys. Rev. A \textbf{66} (2002) 044305.

\bibitem{wang-sun}
X. Wang, H.-B. Li, Z. Sun, Y.-Q. Li, J. Phys. A: Math. Gen. \textbf{38} (2005) 8703.

\bibitem{su}
X. Q. Su, A. M. Wang, Phys. Lett. A \textbf{369} (2007) 196.

\bibitem{markham}
D. Markham, J. Anders, V. Vedral, M. Murao, A. Miyake, EPL \textbf{81} (2008) 40006.

\bibitem{nakata}
Y. Nakata, D. Markham, M. Murao, Phys. Rev. A \textbf{79} (2009) 042313.

\bibitem{rice}
E. Dagotto, T. M. Rice, Science \textbf{271} (1996) 618.

\bibitem{dagotto}
E. Dagotto, Rep. Prog. Phys. \textbf{62} (1999) 1525.

\bibitem{gelfand}
M. P. Gelfand, Phys. Rev. B \textbf{43} (1991) 8644.

\bibitem{richter}
O. Derzhko, J. Richter, Eur. Phys. J. B \textbf{52} (2006) 23.

\bibitem{schmidt}
O. Derzhko, J. Richter, A. Honecker, H.-J. Schmidt, Low Temp. Phys. \textbf{33} (2007) 745.

\bibitem{derzhko}
O. Derzhko, T. Krokhmalskii, J. Richter, Phys. Rev. B \textbf{82} (2010) 214412.

\bibitem{takano}
K. Takano, J. Phys. A: Math. Gen. \textbf{27} (1994) L269.

\bibitem{mila}
F. Mila, Eur. Phys. J. B \textbf{6} (1998) 201.

\bibitem{oitmaa}
Z. Weihong, V. Kotov, J. Oitmaa, Phys. Rev. B \textbf{57}(1998) 11439.

\bibitem{honecker}
A. Honecker, F. Mila, M. Troyer, Eur. Phys. J. B \textbf{15} (2000) 227.

\bibitem{high-t}
O. Rojas, E. V. C. Silva, S. M. de Souza, M. T. Thomaz, Phys. Rev. B \textbf{69} (2004) 134405.

\bibitem{shik}
H. Y. Shik, R. Chen, Phys. China, \textbf{5} (2010) 188.

\bibitem{michaud}
F. Michaud, T. Coletta, S. R. Manmana, J.-D. Picon, F. Mila, Phys. Rev. B \textbf{81} (2010) 014407.

\bibitem{ladder}
T. Verkholyak, J. Stre\v{c}ka, J. Phys. A: Math. Theor. \textbf{45} (2012) 305001.

\bibitem{taras}
T. Verkholyak, J. Stre\v{c}ka, Condens. Matter Phys. (2013) in press.

\bibitem{diana}
D. Antonosyan, S. Bellucci, V. Ohanyan, Phys. Rev. B \textbf{79} (2009) 014432.  

\bibitem{vadim}
V. Ohanyan, Phys. Atom. Nucl. \textbf{73} (2010) 494.

\bibitem{yan-qi} 
Yan Qi, An Du, Yan Ma, Phys. Lett. A \textbf{377} (2012) 27.

\bibitem{niggemann}
H. Niggemann, G. Uiminy, J. Zittartz, J. Phys.: Condens. Matter \textbf{9} (1997) 9031; \textbf{10} (1998) 5217.

\bibitem{baxter}
R. J. Baxter, Exactly Solved Models in Statistical Mechanics (Academic Press, New York, 1982).

\bibitem{wooters}
W. K. Wooters, Phys. Rev. Lett. \textbf{80} (1998) 2245.

\bibitem{amico}
L. Amico, A. Osterloh, F. Plastina, R. Fazio, Phys. Rev. A \textbf{69} (2004) 022304.

\bibitem{bukman}
D. J. Bukman G. An, J. M. J. van Leeuwen, Phys. Rev. B \textbf{43} (1991) 13352. 

\bibitem{spra}
O. Rojas, M. Rojas, N. S. Ananikyan, S. M. Souza, Phys. Rev. A \textbf{86} (2012) 042330.

\end{thebibliography}
\end{document}